\documentclass[nofootinbib,twocolumn]{revtex4}
\usepackage{graphicx}
\usepackage{amssymb}
\usepackage{amsmath}
\usepackage{bm}
\usepackage{hyperref}
\begin{document}

\title{Degeneracy between warm and coupled cold dark matter: A clarifying note}

\author{Hermano Velten$^{1,2}$}\email{velten@pq.cnpq.br}
\author{Humberto Borges$^{3,4}$}\email{humberto@ufba.br}
\author{Thiago R. P. Caram\^es$^4$}\email{thiago.carames@port.ac.uk}

\affiliation{$^1$Departamento de F\'isica, Universidade Federal do Esp\'{\i}rito Santo, Campus Goiabeiras, Vit\'oria, 29075-910 Brazil}
\affiliation{$^2$CPT, Aix Marseille Universit\'e, UMR 7332, 13288 Marseille,  France}
\affiliation{$^3$Instituto de F\'isica, Universidade Federal da Bahia
Campus de Ondina, Salvador, 40210-340 Brazil}
\affiliation{$^4$Institute of Cosmology \& Gravitation, University of Portsmouth, Dennis Sciama Building, Burnaby Road, Portsmouth, PO1 3FX, United Kingdom}

\begin{abstract} Wei et al [PRD 88, 043510 (2013)] have proposed the existence of a cosmological degeneracy between warm dark matter (WDM), modified gravity and coupled cold dark matter (CDM) cosmologies at both the background expansion and the growth of density perturbation levels, i.e., corresponding cosmological data would not be able to differentiate such scenarios. Here, we will focus on the specific indistinguishability between a warm dark matter plus cosmological constant ($\Lambda$) and coupled scalar field-CDM scenarios. Although the statement of Wei et al is true for very specific conditions we present a more complete discussion on this issue and show in more detail that these models are indeed distinguishable. We show that the degeneracy breaks down since coupled models leave a specific signature in the redshift space distortion data which is absent in the uncoupled warm dark matter cosmologies. Furthermore, we complement our claim by providing the reasons which suggest that even at nonlinear level a breaking of such apparent equivalence is also expected.\\

\textbf{Key-words}: Cosmology, Dark Matter, Dark Energy.

PACS numbers: 98.80.-k, 95.36.+x, 95.35.+d, 04.50.-h
\end{abstract}

\maketitle

\section{Introduction}

The actual nature of the dark matter particle remains unknown but the standard cosmological model relies on a candidate belonging to the WIMP (weakly interacting massive particle) category \cite{WIMP}. In this acronynm ``massive'' means such particles are heavy, with masses $O(GeV)$, i.e., they were non-relativistic at freeze-out. As a consequence, since this ensemble of cold dark matter (CDM) particles presents a very low velocity dispersion the pressure of this component is assumed to be irrelevant for the background expansion, i.e., $p_{c}=0$. 

Lighter particles $m \sim O(keV)$ would have decoupled still in the non-relativistic regime and represent the basis of the Warm Dark Matter (WDM) \cite{WDM}. Consequently, their relativistic free-streaming is able to prevent structure formation below the dwarf galaxy scale and since they have a non-negligible thermal velocity there exists a positive pressure contribution. Such pressure can play a role at both the background expansion level and the perturbative sector since it leads to a effective speed of sound. This contribution is sometimes parametrized by a constant equation of state parameter $w_{w}=p_{w}/\rho_{w}$\footnote{In our notation, the subscript $c$ refers to CDM while the subscript $w$ to WDM.}. Current constraints limit the darm matter equation of state to the range $\left|w_{w} \right|\lesssim 10^{-3}$ values \cite{EoSDM}. 

The prevailing view for the dark matter sector also states that these particles should interact weakly. This means that they have very small cross-sections in scattering processes. However, on large scales, dark matter could eventually transfer energy-momentum to the dark energy sector whose nature is still not known. The simplest DE candidate is a cosmological constant, but barotropic fluids and dynamical fields have been also associated to the accelerated cosmic expansion. The suggestion of a possible energy-momentum transfer between DM and DE gave rise to interacting/coupled cosmological models \cite{Int} (see a recent review in \cite{IntReview}). In this scenario, the densities of DM and DE are somehow connected leading to a possible way out for the cosmic coincidence problem (see \cite{rodrigo} and references therein).

Some cosmological observations are capable to probe the expansion rate of the universe $H$. A remarkable example is the Supernovae type Ia data. From Friedmann's equation however we know that $H$ is determined by the sum of all cosmic components. Then, we are not able to distinguish the nature of each single cosmic fluid or different cosmologies which lead to the same $H$. Moreover, a similar situation can also happen when studying the evolution of first order matter density perturbations which give us a direct understanding of how structures grow. If some set of observations are indeed unable to distinguish the actual physics behind that observable then we are left with the so called cosmic degeneracy problem (see for example related discussion in Refs. \cite{kunz, avilez, saulo}).

Wei et al \cite{wei} have proposed a possible equivalence between three distinct cosmological models:

\begin{itemize}
\item A) WDM which does not interact to dark energy in the form of a cosmological constant $\Lambda$;
\item B) CDM coupled to a scalar dark energy field;
\item C) A CDM model driven by a modified gravity theory parametrized in terms of an effective gravitational coupling $G_{eff}$.
\end{itemize}

In comparison to the minimal standard $\Lambda$CDM cosmology, both models have one additional degree of freedom. Namely, model A has the constant equation of state parameter of WDM $w_{w}$, model B has the coupling $Q$ between dark matter and the scalar field dark energy, and model C has the constant effective gravitational coupling $G_{eff}$.
Ref. \cite{wei} argues that by setting the same background expansion $H$ and the same density contrast $\delta$ in such models, then one can find a exact mapping between quantities in both models. Therefore, once one satisfies such mapping, cosmological data for the background expansion and the growth history would not be able to differentiate specific signatures of either WDM particles, dark matter-dark energy coupling or modified gravity. Such claim has also motivated further investigations \cite{citewei}.

We do not focus here on the possible degeneracy involving modified gravity scenarios since this discussion also relies on screening mechanisms to recover the success of general relativity in local tests. In any case, we follow recent literature and we support the ideia that modified gravity models can be distinguishable from other cosmologies, more specifically, interacting cosmologies \cite{MGsign,koyamaMGinter}.

The purpose of this work is twofold: 1) to extend the analysis of Ref. \cite{wei} and to enrich the discussion on the degeneracy between $\Lambda$WDM and coupled CDM scenarios and 2) to identify specific cosmological signatures which are able to break such degeneracy. We anticipate that the proposed degeneracy breaks down at linear and non-linear levels.

In the next section we review the basic dynamics presented in Ref. \cite{wei} as long as their main outcomes. In section III we present in each subsection one argument advocating that the models A and B can be in fact distinguishable. In section IV, we also provide an analysis concerning the non-linear clustering to reinforce our viewpoint. We conclude in the final section.

\section{Warm dark matter and coupled dark matter: Dynamics}

Let us introduce a more general dynamics associated to model A in which the universe contains only WDM and dark energy. Hence, the Friedmann equation reads
\begin{equation}
3H^2=\kappa^2 (\rho_{w} + \rho_{de}),
\end{equation}
where $\kappa^2=8\pi G$. Since there is no interaction between such components
\begin{eqnarray}
\dot{\rho}_w+3H(\rho_{w}+p_{w})&=&0\\
\dot{\rho}_{de}+3H(\rho_{de}+p_{de})&=&0,
\end{eqnarray}
where the equation of state parameter of a fluid $w = p / \rho$ can even be a function of time. However, in standard cosmology one has for CDM $w_{c}=0$ while for relativistic fluids $w_{r}=1/3$. Restricting to model A, $w_{w}$ will be a free constant and $w_{de}=-1$.

At first order, sub-horizon density perturbations $\delta=\delta \rho / \rho$ in the matter fluid obeys to
\begin{eqnarray}\label{deltawdm}
\delta^{\prime \prime}_w + \left[2-3(2w_{w}-c^2_{ad})+\frac{H^{\prime}}{H}\right]\delta^{\prime}_w\\ \nonumber
= \frac{3}{2}\Omega_{w} \delta_w \left(1-6c^2_{ad} + 8 w_{w} -3 w^2_{w}+\frac{k^2 c^2_{eff}}{a^2 H^2}\right).
\end{eqnarray}
where the symbol $^{\prime}$ (prime) means derivative with respect to ln $a$. The effective speed of sound has been defined as $c^2_{eff}=\delta p / \delta \rho$. For adiabatic fluids this quantity reads $c^2_{eff}=c^2_{ad}= p^{\prime}/\rho^{\prime}$. In model A, $c^2_{eff}=w_w$. We have also defined the fractionary energy density such as $\Omega_{w}=\kappa^2 \rho_{w} / 3H^2$.

Now we describe model B with more details. For the scalar field dynamics $\phi$, its energy density and pressure are, respectively,
\begin{equation}
\hat{\rho}_{\phi}=\frac{\dot{\phi}^2}{2}+V(\phi) \hspace{0.4cm}{\rm and}\hspace{0.4cm} 
\hat{p}_{\phi}=\frac{\dot{\phi}^2}{2}-V(\phi), 
\end{equation}
where the symbol `` $\hat{}$ '' (hat) refers to quantities belonging to model B. A specific scalar field dynamics corresponds to a choice of the potential $V(\phi)$. Then, the corresponding Friedmann's equations reads
\begin{equation}
3\hat{H}^2=\kappa^2 (\hat{\rho}_c+\hat{\rho}_{\phi}).
\end{equation}
An interaction between dark matter and dark energy corresponds to a source terms in their energy conservation equations. More specifically, and for the purposes of our work, we present here the coupling adopted by Amendola \cite{amendola} which was used in \cite{wei} 
\begin{eqnarray}
\dot{\hat{\rho}}_c+3H(\hat{\rho}_c+p_c)&=&-\kappa Q \hat{\rho}_c\dot{\phi},\\
\dot{\hat{\rho}}_{\phi}+3H(\hat{\rho}_{\phi}+p_{\phi})&=&\kappa Q \hat{\rho}_c\dot{\phi}.
\end{eqnarray}
The new degree of freedom of the model is encoded in the dimensionless arbitrary function $Q\equiv Q(\phi)$.

The dynamics of the coupled scalar field to CDM is therefore determined by
\begin{equation}
\ddot{\phi}+3\hat{H}\dot{\hat{\phi}}+\frac{d V}{d\phi}=\kappa Q \hat{\rho}_c.
\end{equation}

For the sub-horizon ($k/H \gg 1$) perturbative dynamics of the coupled CDM component we have \cite{amendolatv}
\begin{equation}\label{deltacoupled}
\hat{\delta}^{\prime\prime}_c+\left(2+\frac{\hat{H}^{\prime}}{H}-\kappa Q \phi^{\prime}\right)\hat{\delta}^{\prime}_c=\frac{3}{2}(1+2Q^2)\hat{\Omega}_c\hat{\delta}_c.
\end{equation}

Let us review the results of Ref. \cite{wei} which are related to the equivalence between a $\Lambda$WDM scenario and the coupled CDM described in the last section. The $\Lambda$WDM model has been assumed to possess the parameters $\Omega_{w0}=0.28$, $w_{w}=const$ and, of course, $w_{de}=-1$. 

The main argument of Wei et al is that by demanding the same background expansion and structure growth history, i.e.,
\begin{equation} 
H=\hat{H}\, \hspace{1cm} {\rm and} \hspace{1cm}\delta_w=\hat{\delta}_c,
\end{equation}
there is a exact mapping between the quantities in both models. From adopting that $H=\hat{H}$ they found
\begin{eqnarray}
(\kappa \phi^{\prime})^2 &=& -3 \hat{\Omega}_c-2\frac{\hat{H}^\prime}{\hat{H}}\\
\kappa Q \phi^{\prime}&=&-3-2 \frac{\hat{H}^{\prime}}{\hat{H}}-\frac{\hat{\Omega}^{\prime}_c}{\hat{\Omega}_c},
\end{eqnarray}
which combined produce an expression for $Q$. The function $\hat{\Omega}_c$ is not yet know and should be in general different from the WDM density $\Omega_{w}$. Then, by assuming $\delta=\hat{\delta}_c$ it is possible to find a first order differential equation for $\hat{\Omega}_{c}$ (Eq. (32) of \cite{wei}). According to Ref. \cite{wei}, by comparing (\ref{deltawdm}) and (\ref{deltacoupled}), and taking the sub-horizon limit ($k/H\gg 1$) of the the former, one finds  

\begin{eqnarray} \label{Omegahat}
&&\left[3(2 w_w -c^2_{ad})+3+2 \frac{\hat{H}^{\prime}}{\hat{H}}+\frac{\hat{\Omega}^{\prime}_c}{\hat{\Omega}_c}\right] \delta^{\prime}_w =\\ \nonumber
&&\frac{3}{2}\delta_w \left[\hat{\Omega}_c(1+2Q^2)-\Omega_w (1-6c^2_{ad}+8w_w-3w^2_w)\right].
\end{eqnarray}

Notice that if $w_w$ and $\Omega_{w0}$ are specified we can obtain $\delta_w$ from Eq. (\ref{deltawdm}) and therefore $\hat{\Omega}_c$ is found after integration of the above equation.

Here we point out a drawback in the analysis of Ref. \cite{wei}. Since WDM has a non-negligible $c^2_{eff}$ the $k^2$-dependence in Eq. (\ref{deltawdm}) can not be eliminated as it was claimed. Equation (\ref{Omegahat}) is therefore formally wrong because of the missing $c^2_{eff} k^2$ term which should be indeed present in the sub-horizon limit of (\ref{deltawdm}) .
 
\section{Distinguishing coupled quintessence-CDM from $\Lambda$WDM}

We have show in the last section that the determination of $\hat{\Omega}_c$ according to (\ref{Omegahat}) is inconsistent within the sub-horizon limit. However, let us assume that the freedom in choosing $w_{w}$ allows one to make it as small as necessary in order to make $c^2_{eff} k^2/a^2H^2 \ll 1$ for any wavenumber, i.e., taking the CDM limit. With this we are temporarily relying that $\hat{\Omega}_c$ is well determined from (\ref{Omegahat}).

In the next subsections we will discuss how to distinguish models A and B even 
keeping this ``mislead'' assumption. 

\subsection{Implications of $\Omega_w \neq \hat{\Omega}_c$}

Part of the statistics of the matter field clustering is encoded in the power spectrum $P(k)= \delta^2_k$. One observable which is affected by the evolution of the matter density is the turnover scale $k_{eq}$ in the matter power spectrum. This scale is set by the equality time between matter and radiation, i.e, when $\Omega_{r}(z_{eq})=\Omega_{m}(z_{eq})$. For the standard $\Lambda$CDM this happens around $z_{eq}\sim 3200$. Modes which enter the horizon during the radiation epoch can not grow due to the Hubble drag caused by the radiative background. Only after $z_{eq}$, when the background is dominated by the matter component, perturbations grow according to $\delta \sim a$ (this is also know as the Meszaros effect). If galaxy surveys are able to probe very large volumes then $k_{eq}$ can be used to constrain cosmological models. Unfortunately, we have so far only moderate bounds on the turnover scale provided by the WiggleZ survey $k_{eq}=0.0160^{+0.0035}_{-0.0041} h$ Mpc$^{-1}$ \cite{wigZ}.

Therefore, since we have in general $\Omega_w \neq \hat{\Omega}_c$ we expect that the $k_{eq}$ will differ in both models.

In the specific example given in \cite{wei} $\hat{\Omega}_c$ is found from integration of Eq. (\ref{Omegahat}) since some arbitrary initial condition for $\hat{\Omega}_c$ is provided. We will also show here that this choice play a crucial role.

Let us assume two sub-models for the coupled scenario,

\begin{itemize}
\item Model AI: $\hat{\Omega}_c (z_i)=0.995$ at a redshift $z_i=1000$;
\item Model AII: $\hat{\Omega}_c (z_i)=0.28$ at a redshift $z_i=0$.
\end{itemize}

Reference \cite{wei} adopted model AI while we model AII should be---see (\ref{deltacoupled})---even more similar to the $\Lambda$WDM scenario (model B).

In model AI we find $\hat{\Omega}_{c0}=0.16$. During this procedure it has also been assumed $h=0.7$ and $w_{w}=0.003$ for the WDM. We have estimated the corresponding turnover scale in this specific coupled CDM example as $k_{eq}=0.0145 h$ Mpc$^{-1}$ ($z_{eq}=3265$). The latter result is the same for AI and AII. On the other hand, the supposed degenerated $\Lambda$WDM cosmology presents $k_{eq}=0.0148 h$ Mpc$^{-1}$ ($z_{eq}=3180$). Although we have shown that there is indeed a difference between the models, it is unfortunately very tiny in light of current precision in determining the turnover scale.

\subsection{Matter power spectrum on small scales}

Since the free-streaming of WDM particles with masses $\sim$keV leads to a cut-off in $P(k)$ on small scales one can wonder whether the absence such feature in CDM coupled models could be used to distinguish models A and B.
In practice, this suppression is implemented via the power spectrum definition $P_{w} (k)=T^2_{lin}(k) P_c (k)$ where the suppression effects are encoded in the {\it ad hoc} function $T^2_{lin}(k)$ which in fact also depends on the WDM particle mass $m_{w}$ \cite{VielT2}. Current constraints from Lyman-$\alpha$ observations place a bound $m_{\rm WDM} \geq 3.3$ keV ($2\sigma$) \cite{wdm33}. Moreover, the WDM signature on the HI 21cm power spectrum measured in the future by SKA \cite{SKA} has been discussed in \cite{WDM21cm}.

Now, let us provide a qualitative comparison between the $\Lambda$WDM and coupled CDM cosmologies concerning the $P(k)$ construction. The matter power spectrum in coupled CDM cosmologies has been also widely studied assuming many different coupling functions \cite{PkCoupled}. One remarkable fact is that one does not observe a sharp suppression on small scales as in WDM case. This already provides a hint to differentiating models A and B.

Another important remark here is related to the numerical simulations outcomes. In general, the standard $\Lambda$CDM model is plagued with the prediction of the large number of satellites and the cusp density profile of halos. In this sense, successful cosmologies should avoid the same CDM clustering patterns. Numerical results for coupled cosmologies usually identify specific signatures which are absent in uncoupled cases  \cite{CoupledSimu1, CoupledSimu2}. But in general, both coupled cosmologies and WDM seem to converge to similar predictions. However, available results for numerical simulations do not allow us the make clear statements on such distinguishability since different research groups have their own numerical suites which differ in aspects like initial conditions and merging assumptions, for instance. So far we are aware of, a direct comparison between coupled CDM and $\Lambda$WDM cosmologies has not yet been performed by the same group. Then, this strategy would lead to a clear picture on the indistinguishability of such models.

\subsection{The redshift space distortions signature of coupled/interacting models}

We calculate here in details the predictions for the redshift space distortions (RSD) in models A and B. In measuring the RSD the key physical quantity is the galaxies peculiar velocities $\vec{v}$ which quantifies how anisotropic the galactic clustering is. The derived quantity $(f \sigma_8)$ which has been widely used for constraining cosmologies captures information from both the expansion growth rate via the definition $f =d {\rm ln} \delta / d {\rm ln}a$ as well as the variance of the density field $\sigma_8$.

In the $\Lambda$WDM scenario we can project the WDM energy momentum tensor $T^{\mu \nu}_{w ;\mu}=0$ parallel to its four-velocity $u^{\mu}_{w}$ such that
\begin{equation}
aH \delta^{\prime}_w+(1+w)\Theta=0
\end{equation}
where $\Theta/aH = \nabla \cdot \vec{v}$. Therefore, we find that
\begin{equation}\label{RSDw}
-\frac{\Theta}{aH}=\left(\frac{f}{1+w}\right)\delta_w.
\end{equation}

For the coupled cosmology, however, we should start from 
\begin{equation}
T^{\mu \nu}_{c ; \mu}=Q^{\nu}.
\end{equation}
The projection of the above equation into $u^{\mu}_{c}$ leads to
\begin{equation}
a \hat{H} \hat{\delta}^{\prime}_c +\frac{a \hat{Q}}{\hat{\rho}_{c}}\delta_c +\hat{\Theta}=0.
\end{equation}
The corresponding velocity divergence in coupled CDM models will be then \cite{WandsRSDCoupled}
\begin{equation}\label{RSDcoupled1}
-\frac{\hat{\Theta}}{a\hat{H}}=\left(\hat{f} + \frac{Q}{\hat{\rho}_c \hat{H}}\right)\hat{\delta}_c.
\end{equation}

It is worth noting that the condition $\delta=\hat{\delta}$ means $f=\hat{f}$ and therefore (\ref{RSDcoupled1}) provides a clear distinguished prediction for the RSD measurements than the $\Lambda$WDM case. Clearly, we have $\hat{\Theta}/\Theta \neq 1$ and therefore the degeneracy is broken at linear level.

\begin{figure}[t]
\includegraphics[width=0.44\textwidth]{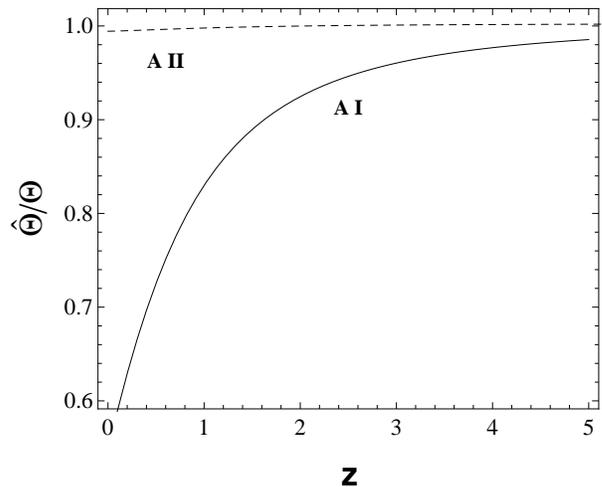}\\ 
\caption{Ratio of the RSD predictions between the coupled CDM and the $\Lambda$WDM models (\ref{RSDratio}). }
\label{fig1}
\end{figure}

In order to quantity whether or not the interacting term leads to a sizable effect in the RSD measurements we calculate the ratio between Eqs. (\ref{RSDcoupled1}) and (\ref{RSDw}),
\begin{equation}\label{RSDratio}
\frac{\hat{\Theta}}{\Theta}=\left(1-\frac{\kappa Q \phi^{\prime}}{H^2 f}\right) (1+w_{w}),
\end{equation}
and plot it in Fig. \ref{fig1}. This shows that the configuration adopted by Wei et al. (AI) leads indeed to a large discrepancy in comparison to the $\Lambda$WDM model. However, the magnitude of this effect depends on the choice of the integration constant for $\hat{\Omega}_c$ since the dashed curve (AII) leads to differences of order $\lesssim 1\%$. Such sensitivity is not reached by RSD surveys and therefore in practice the degeneracy can persist.

\section{Distinguishability at the non-linear level}\label{secnonlinear}
Differences between WDM and CDM during the final process of clustering where effects like dynamical friction, tidal stripping, and tidal heating takes place are expected \cite{NONWDMSIMU}, but such details should be investigated via numerical techniques. 

The consequences at nonlinear level of possible interactions in the dark sector have been received considereable attention from cosmologists in the last years \cite{baldi,casas}. In these works the authors analyse the impacts of the interacting dark energy on small scales, gravitational bias between baryons and cold dark matter, halo mass function and the full nonlinear matter power spectrum obtained from simulations, comparing with a fiducial $\Lambda$CDM framework. In this context, recently some authors succeeded to find new fitting functions that reproduce the nonlinear power spectrum provided by the N-body simulations \cite{casas}. Likewise, similar investigations looking for a potential signature of a WDM model on the nonlinear power spectrum evolution and other nonlinearity effects can be found in \cite{VielT2,wdm33}.

In order to reinforce our claim we extend our investigation to the nonlinear ambit, by adopting the spherical collapse mechanism. Let us notice that our line of reasoning had so far the Fourier space as standpoint, thus a glance at the nonlinear regime enriches our analysis allowing us to check the distinguishability of the both models also in the real space. For sake of simplicty we work within a neo-newtonian fluid description and consider the top-hat profile for the spherical overdensity. 
In models satisfying this condition the perturbed physical quantities are assumed to be homogeneous throughout the collapse, so that the gradients both of the density and pressure perturbations are neglected. Our aim is to show that also within this framework the equivalence between $\Lambda$WDM and coupled $\phi$CDM models claimed by Wei {\it et al} is not fulfilled. 
So let us now look at the dynamical equations describing the time evolution of the matter perturbation in these models. For the $\Lambda$WDM model the clustering component is the WDM whose equation of state is constant. In this case the spherical collapse predicts the following perturbative equation \cite{pace}
\begin{eqnarray}
\label{wdm}
&&\ddot{\delta}_w+2H\dot{\delta}_w-\frac{3}{2}\Omega_w(1+3w_{w})(1+w_{w})\delta_w \\&=&\frac{3}{2}\Omega_w(1+3w_{w})(1+w_{w})\delta^2_w+\frac{(4+3w_{w})}{3(1+w_{w})}\frac{\dot{\delta}^2_w}{(1+\delta_w)}\nonumber ,
\end{eqnarray}
whereas for the interacting $\phi$CDM the analogue equation comprises (which can be compared with the equation (17) of the reference \cite{devi})
\begin{eqnarray}
\label{phicdm}
&&\ddot{\hat{\delta}}+(2H-\kappa Q\dot{\phi})\dot{\hat{\delta}}-\left[2H\kappa Q\dot{\phi}+\frac{d}{dt}(\kappa Q\dot{\phi})+\frac{3}{2}H^2\hat{\Omega}_{c}\right]\hat{\delta}\nonumber\\&=&\frac{3}{2}H^2\hat{\Omega}_{c}\hat{\delta}^2+\frac{1}{3(1+\hat{\delta})}\left[4\dot{\hat{\delta}}^2-5\kappa Q\dot{\phi}\dot{\hat{\delta}}\delta+(\kappa Q\dot{\phi})^2\hat{\delta}^2\right]\ ,
\end{eqnarray}

For the purpose of giving the reader a proper idea about the role played by the nonlinear terms in both models we separate them out the linear ones by displaying them at the right side of each corresponding equation. The breaking of the alleged equivalence claimed by Wei {\it et al}, at linear level, becomes even clear when we observe the full nonlinear equations above. By using the condition $\delta=\hat{\delta}$ and $H=\hat{H}$ a straightforward comparison between these dynamical equation is impossible. The nonlinear increments appearing in (\ref{wdm}) are proportional to $\dot{\delta}^2(1+\delta)^{-1}$ and $\delta^2$, whilst in the equation (\ref{phicdm}) 
the nonlinearity effects contribute with the terms $\dot{\hat{\delta}}^2(1+\hat{\delta})^{-1}$, $\hat{\delta}^2$, $\dot{\hat{\delta}}\hat{\delta}(1+\hat{\delta})^{-1}$ and $\hat{\delta}^2(1+\hat{\delta})^{-1}$. It is important to notice that unlike the former two terms the latter two ones don't have any counterparts in the equation (\ref{wdm}). Furthermore these additional terms arise as direct contributions coming from the interaction assumption at nonlinear level, since they vanish automatically when the interaction is switched off ($\phi$=const.), and the $\Lambda$CDM model is recovered. This feature allows us to interpret the terms $\dot{\hat{\delta}}\hat{\delta}(1+\hat{\delta})^{-1}$ and $\hat{\delta}^2(1+\hat{\delta})^{-1}$ in the equation (\ref{phicdm}) as a sign of the existence of the interactions in the dark sector and since it is expected that these contributions become more and more important as the collapse reaches its final stages, we can infer that these terms can leave imprints on the estimate of abundance of collapsed objects which shall be different from the predictions coming from $\Lambda$WDM. Therefore this difference seems to be crucial to prevent the direct mapping between both models.      

In fact it is possible to confirm the non-equivalence between these two scenarios and verify that the spherical collapse is indeed affected diferently in the both models. In order to provide an estimation on the magnitude of such difference we have solved numerically Eqs. (\ref{wdm}) and (\ref{phicdm}) adopting model AII for comparison. 
Using the same initial condition for both overdensities we found that $\delta^{NL}_{w}(z=0)$ and $\hat{\delta}^{NL}_{c} (z=0)$ differ at the $22\%$ level.
For sake of comparison, from Fig. 1 we can infer that $\hat{\Theta}$ and $\Theta$ calculated at $z=0$ differ only at the $1 \%$ level. Therefore, the nonlinear analysis seems to be more effective in differentiating these cosmologies.

As is well known the abundance of collapsed objects in a given cosmological scenario can be appraised using the linear overdensity evaluated at the collapse redshift $\delta_{crit}\equiv\delta_{linear}(z=z_{collapse})$, the so-called {\it threshold}.
The collapse time means the instant at which the nonlinear density enhancement blows up which in general has a strong dependence upon the amplitude of the initial overdensity: the huger (tinier) the initial density contrast is, earlier (later) the collapse occurs. Thus, the values of $\delta^{NL}_{w}(z=0)$ and $\hat{\delta}^{NL}_{c} (z=0)$ reported above could be arbitrarily amplified by adjusting the initial conditions until the collapse (divergent nonlinear overdensity) is reached. 
Thus by aiming to verify how the degeneracy breaking of the models influences the corresponding threshold values, we computed such parameter for each model at a specific time $z_{c}$. We pursued the same redshift collapse for each perturbative equations and adjusted it to $z_{c}\sim0$ by means of a meticulous choice of the initial conditions. For the $\Lambda$WDM model we set $\delta_w(z_i=1000)=3.4467\times10^{-3}$ and $\delta'_w(z_i=1000)=6.43943\times 10^{-8}$, whereas for the coupled model we have $\delta_c(z_i=1000)=3.779\times 10^{-3}$ and $\delta'_c(z_i=1000)=10^{-8}$. The both cases provide redshift collapse values at the order $z_{c}\sim10^{-5}$, and both values are approximately the same differing just in the ninth decimal place. We observed that the discrepancy in the both threshold values at $z_c$ corresponds to a $\sim2.2\%$ level. Although quite small, the distinguishabiliy appears again through a difference between the corresponding $\delta_{crit}$ for each aforesaid scenario. This tell us that investigating the amount of bound objects over a certain range of mass, following for instance the Press-Schechter formalism or Sheth-Tormen fitting, indicates a possible way to disrupt the supposed degeneracy between the both models.

\section{Final Remarks}
We have discussed about the possible indistinguishability at background and linear order between coupled CDM-quintessence and $\Lambda$WDM scenarios presented in \cite{wei}. Such situation is indeed quite intriguing since most of the available observational data would be unable to differentiate cosmological models.

Assuming the strategy of \cite{wei} such that the scale dependence of equation (\ref{deltawdm}) is eliminated then, even so we were able to show that differences between such models persist. In this case the question becomes actually how sizable are such differences and whether or not observations of the structure growth history can distinguish them. 

In particular the expression for the redshift space distortions in the coupled cosmology (\ref{RSDcoupled1}) has a clear signature of the interaction $Q$.

Finally, we have also shown in section \ref{secnonlinear} that the nonlinear density evolution of such models is indeed different.

\textbf{Acknowledgement}: We acknowledge enlightening conversations with Marco Bruni, Federico Piazza, Valentina Salvatelli and David Wands. We thank CNPq for financial support. HV also thanks the A*MIDEX Foundation under Contract ANR-11-IDEX-0001-02 and HB thanks Fapesb.

\end{document}